\documentclass[conference]{IEEEtran}
\usepackage{graphicx}
\usepackage{amsmath}
\usepackage{mathtools}
\usepackage{latexsym}

\usepackage{amsthm}
\theoremstyle{definition}
\newtheorem{theorem}{Definition}
\newtheorem*{theorem*}{Definition}
\newtheorem{definition}[theorem]{Definition}
\newtheorem*{definition*}{Definition}

\usepackage{listings}

\usepackage{algorithm}
\usepackage{algorithmic}

\begin{document}

\title{A new perspective of paramodulation complexity by solving massive 8 puzzles}

\author{

\IEEEauthorblockN{Ruo Ando}
\IEEEauthorblockA{
National Institute of Informatics\\
2-1-2 Hitotsubashi, Chiyoda-ku, Tokyo \\101-8430 Japan\\
}
\and
\IEEEauthorblockN{Yoshiyasu Takefuji}
\IEEEauthorblockA{Keio University\\
5322 Endo Fujisawa, Kanagawa \\252-0882 Japan\\
}
}

\maketitle

\begin{abstract}
A sliding puzzle is a combination puzzle where a player slide pieces along certain routes on a board to reach a certain end-configuration.
In this paper, we propose a novel measurement of complexity of massive sliding puzzles with paramodulation which is an inference method of automated reasoning. 
It turned out that by counting the number of clauses yielded with paramodulation, we can evaluate the difficulty of each puzzle. 
In experiment, we have generated 100 * 8 puzzles which passed the solvability checking by countering inversions. 
By doing this, we can distinguish the complexity of 8 puzzles with the number of generated with paramodulation. 
For example, board [2,3,6,1,7,8,5,4, hole] is the easiest with score 3008 and board [6,5,8,7,4,3,2,1, hole] is the most difficult with score 48653.
Besides, we have succeeded to obverse several layers of complexity (the number of clauses generated) in 100 puzzles. 
We can conclude that proposal method can provide a new perspective of paramodulation complexity concerning sliding block puzzles.
\end{abstract}

\IEEEpeerreviewmaketitle

\section{Introduction}

A sliding puzzle (also called as sliding block puzzle) is a combination puzzle where a player slide pieces along certain routes on a board to reach a certain end-configuration (state). 
The pieces are usually numbered, and sometimes may be imprinted with colors, patterns and sections of a large picture. 


In nature, sliding puzzles are two-dimensional even if the sliding is facilitated by encaged marbles or three-dimensional tokens.

\begin{figure}[ht]
\centering
\includegraphics[scale=0.5]{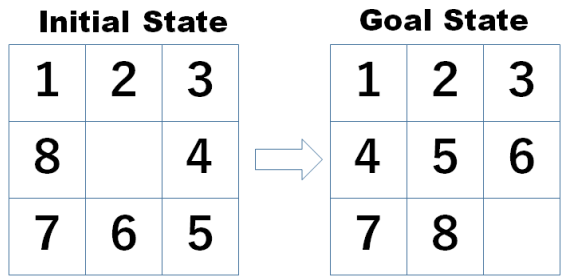}
\caption{Initial state and goal state of 8 puzzle.}
\end{figure}

In sliding puzzles, a player is prohibited to lift any piece off the board. This constraint separates sliding puzzles from rearrangement puzzles. 
Consequently, discovering routes opened up by each move with the two-dimensional confines of the board is interesting point of solving sliding block puzzles.

Figure 1 shows the example of sliding puzzle. The puzzle has 9 square slots on a square board.
The first 8 slots have square pieces. The 9th slot is empty. 

Historically, Noyes Chapman invented the oldest type of sliding puzzle which is the fifteen puzzle in 1880.
Folklore tells us that in 1886, puzzle master Sam Loyd offered a one-thousand dollar prize if anyone could swap tile 14 and 15 and return the other tiles to their original slots.

Sliding block can be represented as the permutation.
A permutation of a set S is a bijection from S onto itself. If the set we permuting is $ A = {1,2, ..., n} $, it is often convenient 
to represent a permutation $ \sigma $ as follows:

\begin{eqnarray}
\sigma = \left\{
\begin{array}{lllll}
1,&2,&3,& ... &\\
\sigma(1),&\sigma(2),&\sigma(3),& ... ,& \sigma(n)
\end{array}
\right\}
\end{eqnarray}

For instance, consider the set A = {1,2,3,4,5,6}. Then the permutation $\pi$,

\begin{eqnarray}
\pi = \left\{
\begin{array}{llllll}
1,&2,&3,&4,&5,&6\\
4,&1,&5,&2,&3,&6
\end{array}
\right\}
\end{eqnarray}

sends 1 to 4, 2 to 1, 3 to 5 and fixes, or leaves unchanged, the element 6.

\section{OTTER}

\subsection{OTTER and its clause sets}
The theorem prover OTTER (Organized Techniques for Theorem-proving and Effective Research) has been developed by W. McCune as a product of Argonne National Laboratory.
OTTER is based on earlier work by E. Lusk, R. Overbeek and others \cite{Lusk2}.
By the research efforts of \cite{McCune1} \cite{McCune2} \cite{Lusk} for certain classes of problem, 
OTTER is widely regarded as the most powerful automated deduction system. 
OTTER adopts the given-clause algorithm and implements the set of support strategy \cite{Wos2}.

In given-clause algorithm, all retained clauses are divided into two sets.
The first sets are called as the set of support (SoS). OTTER starts with the retention of a set of support including all of the choses input
clauses. During the run, the initial set of support and the clauses which are generated are retained.
The second set is the usable list. At beginning phase of reasoning, the usable list is not included into the initial set of support. 
The usable list are the clauses which were once in the set of support but have been already picked up as the focus of attention for deducing additional clauses.
More technically, in detail, OTTER maintains four lists of clauses in reasoning process.

\begin{enumerate}
\item Usable. This list works as a rule by keeping clauses which are available to make inferences.

\item SoS. Clauses is regarded as facts. Set of support are not used to make inferences. They are kept to participate in the search.

\item Passive. They are specified to be used only for forward subsumption and unit conflict. The passive list does not participate in the search.
The passive list does not change from the start of reasoning process as fixed input.

\item Demodulators. Demodulators are used to rewrite newly inferred clauses with equalities.
\end{enumerate}

In this paper, particularly, we focus on the size of set of support list. 
Set of support is important indicator for introspecting the reasoning process. 

\subsection{Given Clause algorithm}

OTTER adopts given-clause algorithm in which the program attempts to use any and all combinations from axioms in given clause. 
In other words, the combinations of clause are generated from given clauses which has been focused on.

\begin{algorithm}
\caption{Given clause algorithm}
\label{alg1}
\begin{algorithmic}[1]
\renewcommand{\algorithmicrequire}{\textbf{Input:}}
\renewcommand{\algorithmicensure}{\textbf{Output:}}
\REQUIRE SOS, Usable List
\ENSURE Proof
\WHILE{until SoS is empty}
\STATE choose a given clause G from SoS;
\STATE move the clause g to Usable List;
\WHILE { c\_1, ..., c\_n in Usable List} 
\WHILE{$ R(c_1,..c_i, G ,c_{i+1},..c_n) exists $}
\STATE $ A \Leftarrow R(c_1,..c_i, G, c_{i+1},..c_n); $
\IF{A is the goal}
\STATE report the proof;
\STATE stop
\ELSE[A is new odd]
\STATE add A to SoS X
\ENDIF
\ENDWHILE
\ENDWHILE
\ENDWHILE
\end{algorithmic}
\end{algorithm}

At line 2, given clause G is extracted from SoS (Set of Support). 
Line 4 and 5 is a loop to use any and all combinations of given clause and Usable List.
In detail, \cite{Slaney} \cite{Graf} discuss the basic framework of given clause algorithm.
To put it simply, given clause algorithm consists of the following steps.

\begin{enumerate}
\item Pick up a clause (called the given clause) from the set of support.
\item Add the given clause to the usable list.
\item Applying the inference rule or rules in the effect, infer all clauses which are generated from the given clause (one parent) and the usable list (other parents). 
\item Process newly inferred clause. 
\item Append each inferred new clause to the SoS. These clause is not discarded as a result of processing. Exactly, this is done in the course of processing the newly generated clause.
\end{enumerate}

In a nutshell, the reasoning program chooses a clause from the clauses which is focused on in the set of support.
The selected clause is called as focal clause or given clause.

\begin{definition}
\underline{Definition of given clause.}
The reasoning program chooses a clause on which to focus from among those in the set of support, where the choice is based 
on various criteria such as the weight of the clause. The chosen clause is based on various criteria from among those
in the clause. The chosen clause is called the ``focal clause'' (formerly the ``given clause''). The algorithm under discussion 
permits the focal clause to be considered by whatever inference rules are being used, where the remaining clauses required by the inference 
rule is selected from the usable list, but not from the set of support.
An equality literal is a literal whose predicate is to be interpreted as meaning ``equal''. The inference rule
yields the clause C from the clauses A and B that are assumed to have no variables in common when A contains a positive 
equality literal and B contains a term which unifies with one of the arguments of that equality literal.
\end{definition}

\section{Paramodulation}

Paramodulation is powerful method of equational reasoning. 
It is the method based on resolution refutations which includes the equality. 
In the view of equational reasoning, paramodulation is a generalization of equality substitution. 
For example, if the equality of $ s = t $ and s occurs in the sentence S, paramodulation can replace $s$
in any of the occurrences. 
Also, on the rule of $ s = t $, if $t$ occurs in S, then reasoning program can replace $s$ with $t$.


\begin{eqnarray}
\frac{C \vee A D \vee \neg B }{(C \vee A) \sigma} \; \; \; if \sigma = mgu(A,B) 
\end{eqnarray}

here, mgu (A, B) denotes a most general unifier of A and B, and factoring: 

\begin{eqnarray}
\frac{C \vee A \vee B }{(C \vee A) \sigma} \; \; \; if \sigma = mgu(A,B) 
\end{eqnarray}

\begin{definition}
\underline{Definition of paramodulation.} 
An equality means if a literal whose predicate is to be represented as equal.
The inference rule of paramodulation yields the clause C between the clause A and B which are assumed to have no variables 
in common. Also, if A contains a positive equality literal and B contains a term which unifies with one of the arguments, 
C is yielded by paramodulation. 
\end{definition}

In the equations above (3)(4), Clause A is called the from clause, clause B is called the into clause, and clause C a paramodulant.
Corresponding to the syntax of OTTER, for example, given the following two clauses, from the first into the second

\begin{verbatim}
EQUAL(a,b).
Q(a).
\end{verbatim}

the clause
\begin{verbatim}
Q(b).
\end{verbatim}

is yielded by adopting paramodulation. For a second example, given the following two clauses, from the first into the second

\begin{verbatim}
EQUAL(sum(x,0), x).
P(sum(sum(a,0), b),c).
\end{verbatim}

Paramodulation yields

\begin{verbatim}
P(sum(a,b),c).
\end{verbatim}

as a paramodulant, from 

\begin{verbatim}
Q(g(f(g(x)))).
EQUAL(g(a),b).
\end{verbatim}

the clause

\begin{verbatim}
Q(g(f(b))).
\end{verbatim}
is deducible.

In general, paramodulation is intended to be utilized, along with resolution, for theorem proving in first-order theories with equality.

Concerning the implementation of OTTER, in paramodulation, two parents and a child are processed 
The parent clauses contain the equality applied for the replacement. The parent clauses are divided into two: from parent and from clause. 
If the equality comes from the literal, the side of the equality unifies with the term which is replaced with the from term.
The replaced term is called as the into term. The literal containing the replaced term is also called as the into literal. 
Also the parent containing the replaced term is called as the into parent or into clause.

\section{Methodology}

\subsection{Setting OTTER's rule set}

As we discussed before, the basic inference mechanism of OTTER is based on 
the given-clause algorithm.
Given-clause algorithm can be viewed as a simple implementation of the set of support strategy.
OTTER maintains four lists of clauses: usable, SoS, demodulator and passive. 
In our case, we cope with two kinds of clauses: usable and SoS.


Horizontal sliding from row[i] to row[i+1] is represented as follows.

\begin{verbatim}
list(usable).
EQUAL(l(hole,l(n(x),y)),l(n(x),l(hole,y))).
end_of_list.
\end{verbatim}

\begin{eqnarray}
\sigma = \left\{
\begin{array}{lllll}
1 & hole & 2 & 3 &\\
1 & 2 & hole & 3 &\\
\end{array}
\right\}
\end{eqnarray}

Vertical sliding from row[i] to row[i+4] is represented as follows.

\begin{verbatim}
list(usable).
EQUAL(l(hole,l(x,l(y,l(z,l(u,l(n(w),v)))))),
l(n(w),l(x,l(y,l(z,l(u,l(hole,v))))))).
end_of_list.
\end{verbatim}

\begin{eqnarray}
\sigma = \left\{
\begin{array}{lllllllll}
1 & hole & 2 & 3 & 4 & 5 & 6 & 7 \\
1 & 2 & 3 & 4 & 5 & hole & 6 & 7 \\
\end{array}
\right\}
\end{eqnarray}

\subsection{Checking the solvability of N puzzles}

In general, to check the solvability of N puzzles, the number of inversions of each number of N slots is calculated. 

For example, if we have the board configuration board [2,3,6,1,7,8,5,4, hole]

(5,2,8,4,1,7, hole, 3,6), the number of inversions are as follows:

\begin{enumerate}
\item 2 precedes 1 - 1 inversions
\item 3 precedes 1 - 1 inversion
\item 6 precedes 1, 5, 4 - 3 inversions
\item 1 precedes none - 0 inversions
\item 7 precedes 5, 4 - 2 inversions
\item 8 precedes 5, 4 - 2 inversions
\item 5 precedes 4 - 1 inversions
\item 4 precedes none - 0 inversions
\end{enumerate}

Total inversions 1+1+3+0+2+2+1+0 = 10 (Even Number) So this puzzle configuration is solvable.
On the other hand, it is not possible to solve an instance of 8 puzzle if number of inversions is odd in the input state. 

\begin{algorithm}
\caption{Checking the solvability of N puzzles}
\label{alg1}
\begin{algorithmic}[1]
\renewcommand{\algorithmicrequire}{\textbf{Input:}}
\renewcommand{\algorithmicensure}{\textbf{Output:}}
\REQUIRE Board[$ x_1, x_2, ..., x_n, hole $]
\ENSURE SOLVABLE or UNSOLVABLE
\STATE $ Board[X | XS] = Board[x_1, x_2, ..., x_n, hole] $
\WHILE{ $XS$ in $Board[X l XS] $ is empty}
\FOR {$i$ in $XS$}
\STATE statements..
\IF {($ X \ne XS[i] $)}
\STATE $ counter[i]++$
\ENDIF
\ENDFOR

\ENDWHILE

\STATE $ line = check(Board[...] \subseteq hole) $

\STATE $ sum = 0 $
\FOR {$i$ to $n$}
\STATE $ sum += counter[i]$
\ENDFOR

\IF {($ line + sum \% 2 == 0 $)}
\STATE $ flag =$ SOLVABLE
\ELSE
\STATE $ flag =$ UNSOLVABLE
\ENDIF

\end{algorithmic}
\end{algorithm}

Algorithm 2 shows the procedure for checking the solvability of N puzzles.
At line 2 to 9, the number of inversions of each slots is counted.
These figures are counted up at line 11 to 14. 
Finally, the sum is checked if it is even or odd number at line 15 to 19.

\subsection{Incrementing the number of generated clauses}

The main loop for inferring and processing clauses and searching for a refutation operates
mainly on the lists usable and SoS.

\begin{enumerate}
\item Choose appropriate $given\_clause$ in SoS;
\item Move $given\_clause$ from $list(SoS)$ to $list(usable)$
\item Infer and process new clauses using the inference rules set. 
\item Newly generated clause must have the $given\_clause$. 
\item Do the retention test on new clauses and append those to $list(SoS)$.
\end{enumerate}

Main loop is depicted in Algorithm 3.

\begin{algorithm}
\caption{Incrementing the number of generated clauses}
\label{alg1}
\begin{algorithmic}[1]

\WHILE{given clause is NOT NULL}
\STATE $ index\_lits\_clash(giv\_cl); $
\STATE $ append\_cl(Usable, giv\_cl); $

\IF{$ splitting() $ }
\STATE $ possible\_given\_split(giv\_cl); $
\ENDIF
\STATE{infer\_and\_process(giv\_cl);}
\STATE{giv\_cl = extract\_given\_clause();}
\STATE{track(the\_number\_of\_generated\_clauses);}

\ENDWHILE
\end{algorithmic}
\end{algorithm}

At line 9, the number of generated clauses is incremented.
After line 8 of picking up the clause from set of support, we can record the current size of set of support. 
By doing this, we can obtain the plot with \# puzzles and the number of generated clauses of Y-axis as shown in the next section.

\section{Experimental results}

In experiment, we have generated 100 sliding puzzles with size 8 * 8.
All generated configurations of 8 puzzle are solvable. 
For each puzzle, we have measured the number of generated clauses with the procedures shown in Algorithm 2.
For simplicity, we have generated the configuration of first 8 slots with random integers ranging from 1 to 8 
and fixed 9th slot to hole, as shown in the left side of Table I.

The number of generated clauses with paramodulation ranges from 3008 (2,3,6,1,7,8,5,4, hole) to 468453 (6,5,8,7,4,3,2,1, hole). 
In the view of complexity of reasoning process, the configuration [ (6,5,8,7,4,3,2,1, hole)] is 155.73 times harder to solve than
the configuration [(2,3,6,1,7,8,5,4, hole)]. 

\begin{table}[htbp]
\begin{center}
\caption{Initial board states and the complexities of paramodulation}
\begin{tabular}{{|l|c|}} \hline
Initial state & clauses generated \\ \hline \hline
2,3,6,1,7,8,5,4,hole & 3008 (easiest) \\ \hline 
2,4,3,8,7,6,5,1,hole & 31344 \\ \hline 
2,5,3,8,6,1,7,4,hole & 272413 \\ \hline 
6,5,8,7,4,3,2,1,hole & 468453 (the most difficult) \\ \hline 
\end{tabular}
\end{center}
\end{table}

Figure 2 and 3 show the number of clauses generated with paramodulation which could be described as paramodulation complexity. 
In both graphs, X-axis is the number 8 puzzles. Y-axis is the number of generated clauses.
For yielding Figure 2 and 3, we have generated 100 * 8 puzzles which are solvable as discussed in section IV-B. 
We have observed large variance in Figure 2. 
Besides, Figure 3 depicts the plot sorted by the number of clauses where the puzzle numbers are shuffled. 

\begin{figure}[htbp]
\centering
\includegraphics[scale=0.55]{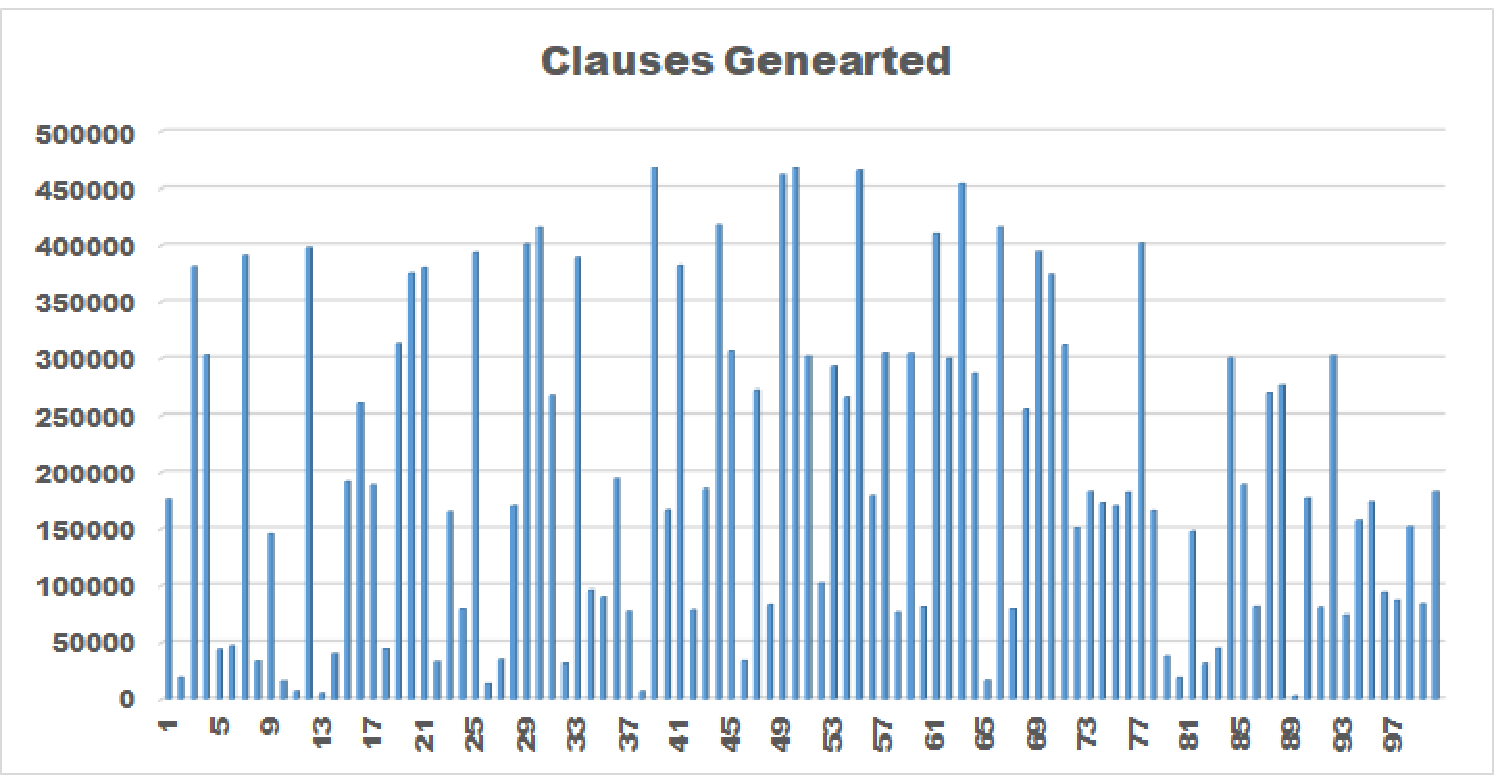}
\caption{Clauses generated by paramodulation. X-axis is the number 8 puzzles. Y-axis is the number of generated clauses.}
\end{figure}

\begin{figure}[htbp]
\centering
\includegraphics[scale=0.55]{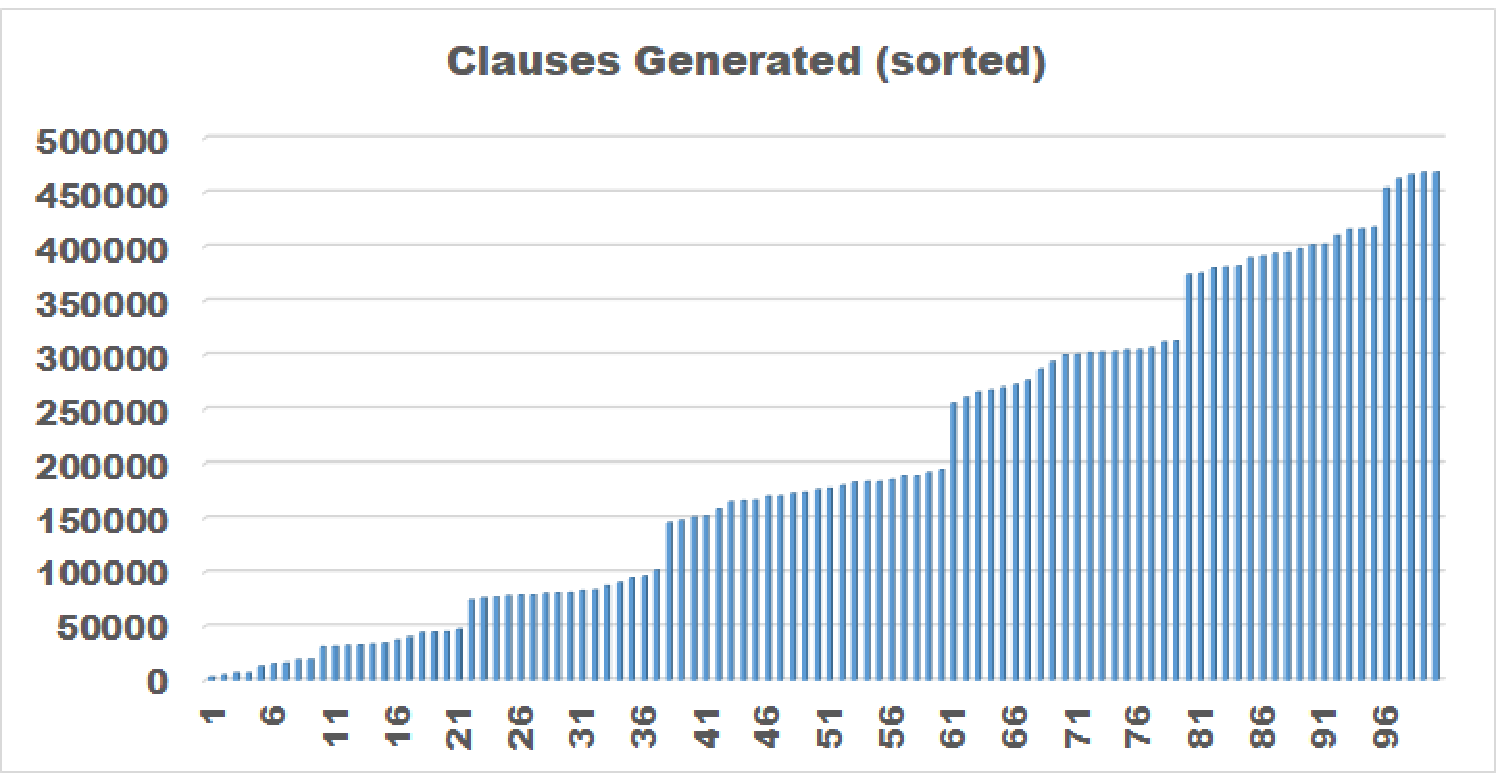}
\caption{Clauses generated by paramodulation. X-axis is the number 8 puzzles. Y-axis is the number of generated clauses.}
\end{figure}

Curiously, in sorted graph in Figure 3, the number of clauses is not increasing linearly. 
Instead, the number of clauses generated with paramodulation is increased drastically around X-axis 22, 38, 62, 79 and 97.
Consequently, we can conclude that there are several layers of complexity in Figure 3.
Also, in each layer, the number of clauses generated is increasing linearly.


\section{Related work}
Archer \cite{Archer} firstly discusses an algorithmic analysis of 15 puzzle. 
In \cite{Archer}, a summary of all possible permutations of slots attained by moving the black block from cell i to cell j effecting the permutation of $ \sigma_i,j $.
Howe \cite{Howe} proposes two approaches in the two kinds of viewpoints: the properties of permutations and graph theory.
Ariyanto \cite{Ariyanto} proposes the new sliding puzzle made with several additional rules from M13 puzzle.
Calabro \cite{Calabro} proposes $ O(n^2) $ time algorithm for deciding the time when tie initial configuration of the n * n puzzle game is solvable.
Conrad \cite{Conrad} discusses 15 puzzle and rubik cube as permutation puzzle.
Bischoff \cite{Bischoff} adopts reinforcement learning to solve 15-puzzle.
Ando \cite{Ando} applies hot list strategy \cite{hotlist} for faster paramodulation-based viral code detection. 
Takefuji proposes the application of paramodulation to translator of Common Lisp \cite{Takefuji}.
Ando and Takefuji applies hot list strategy based on paramodulation for faster graph coloring \cite{Ando2}. 

Paramodulation originated as a development of resolution \cite{robinson}, one of the main computational methods in first-order logic, see \cite{Bachmair}.
For improving resolution-based methods, the study of the equality predicate has been particularly important, since reasoning with equality
is well-known to be of great important of mathematics, logic and computer science. 

Dan Carson and Larry Wos developed a resolution based theorem prover they called P1 which stands for ``Program 1''. P1 is the founder of OTTER and includes basic
strategy of OTTER including the set-of-support strategy \cite{Wos2}, unit preference \cite{Wos3} and paramodulation. 
P1 is the first implementation of theorem prover where Wos's invention of the paramodulation inference rule \cite{robinson2} is experimented.
RW1 which stands for ``Robinson-Wos 1'' which is the product from the collaboration of Wos and George Robinson.
RW1 adopts the paramodulation inference rule as well as demodulation \cite{robinson2}.
Also, RW1 is based on the concept by Knuth and Bendix, who independently formulated paramodulation and demodulation in the view of a complete set of reductions in their 1970 paper \cite{knuth}.

\section{Conclusion}

In this paper we propose a novel method for providing new perspective of paramodulation complexity by solving 100 sliding block puzzles (8 puzzle). 
Paramodulation is designed based on the concept of generalization of a substitution rule for equality. 
We have counted the number of clauses generated with paramodulation as the complexity of each sliding block puzzle as shown in Algorithm 3.
As a result, a wide range of complexity of 100 solvable 8 puzzles have been measured. 
For example, board [2,3,6,1,7,8,5,4, hole] is the easiest with score 3008 and board [6,5,8,7,4,3,2,1, hole] is the most difficult with score 48653.

There have been many research efforts on the measurement and evaluation of the complexity of sliding block puzzles. 
However, the method for coping with complexity of the puzzle in the aspect of the computation cost in automated reasoning has never been proposed. 
We have succeeded to figure out more computational method for the comparison of difficulty of 100 * 8 puzzles with the help of automated reasoning. 
Besides, we have observed several layers of complexity (the number of clauses generated) in 100 puzzles as shown in Figure 3.
We can conclude that proposal method can provide a new perspective of paramodulation complexity concerning sliding block puzzles.

\end{document}